# AN ARCHITECTURE-BASED DEPENDABILITY MODELING FRAMEWORK USING AADL

Ana-Elena Rugina, Karama Kanoun and Mohamed Kaâniche
LAAS-CNRS, University of Toulouse
7 avenue Colonel Roche, 31077 Toulouse Cedex 4, France
Phone:+33(0)5 61 33 62 00, Fax: +33(0)5 61 33 64 11
e-mail: {rugina, kanoun, kaaniche}@laas.fr

**ABSTRACT**

For efficiency reasons, the software system designers' will is to use an integrated set of methods and tools to describe specifications and designs, and also to perform analyses such as dependability, schedulability and performance. AADL (Architecture Analysis and Design Language) has proved to be efficient for software architecture modeling. In addition, AADL was designed to accommodate several types of analyses. This paper presents an iterative dependency-driven approach for dependability modeling using AADL. It is illustrated on a small example. This approach is part of a complete framework that allows the generation of dependability analysis and evaluation models from AADL models to support the analysis of software and system architectures, in critical application domains.

## 1. Introduction

The increasing complexity of software systems raises major concerns in various critical application domains, in particular with respect to the validation and analysis of performance, timing and dependability requirements. Model-driven engineering approaches based on architecture description languages (ADLs) aim at mastering this complexity at the design level. Over the last decade, considerable research has been devoted to ADLs leading to a large number of proposals [1]. In particular, AADL (Architecture Analysis and Design Language) [2] has received an increasing interest from the safety-critical industry (i.e., Honeywell, Rockwell Collins, Lockheed Martin, the European Space Agency, Airbus) during the last years. It has been standardized under the auspices of the International Society of Automotive Engineers (SAE), to support the design and analysis of complex real-time safety-critical applications. AADL provides a standardized textual and graphical notation, for describing architectures with functional interfaces, and for performing various analyses to determine the behavior and performance of the system being modeled. AADL has been designed to be extensible to accommodate analyses that the core language does not support, such as dependability and performance.

In critical application domains, one of the challenges faced by the software engineers concerns: 1) the description of the software architecture and its dynamic behavior taking into account the impact of errors and failures, and 2) the evaluation of quantitative measures of relevant dependability properties such as reliability, availability and safety, allowing them to assess the impact of errors and failures on the service. For pragmatic reasons, the designers using an AADL-based engineering approach are interested in using an integrated set of methods and tools to describe specifications and designs, and to perform dependability evaluations. The AADL Error Model Annex [3] has been defined to complement the description capabilities of the AADL core language standard by providing features with precise semantics to be used for describing dependability-related characteristics in AADL models (faults, failure modes and repair assumptions, error propagations, etc.). AADL and the AADL Error Model Annex are supported by the Open Source AADL Tool Environment (OSATE)[1].

At the current stage, there is a lack of methodologies and guidelines to help the developers, using an AADL based engineering approach, to use the notations defined in the standard for describing complex dependability models reflecting real-life systems with multiple dependencies between components. The objective of this paper is to propose a structured method for AADL dependability model construction. The AADL model is built and validated iteratively, taking into account progressively the dependencies between the components.

The approach proposed in this paper is complementary to other research studies focused on the extension of the AADL language capabilities to support formal verifications and analyses (see e.g. [4]). Also, it is intended to be complementary to other studies focused on the integration of formal verification, dependability and performance related activities in the general context of model driven engineering approaches based on ADLs and on UML (see e.g., [5-9]).

---

[1] http://lwww.aadl.info/OpenSourceAADLToolEnvironment.html

The remainder of the paper is organized as follows. Section 2 presents the AADL concepts that are necessary for understanding our modeling approach. Section 3 gives an overview of our framework for system dependability modeling and evaluation using AADL. Section 4 presents the iterative approach for building the AADL dependability model. Section 5 illustrates some of the concepts of our approach on a small example and section 6 concludes the paper.

## 2. AADL concepts

**The AADL core language** allows analyzing the impact of different architecture choices (such as scheduling policy or redundancy scheme) on a system's properties [10]. An architecture specification in AADL is an hierarchical collection of interacting components (software and compute platform) combined in subsystems. Each AADL component is modeled at two levels: in the component type and in one or more component implementations corresponding to different implementation structures of the component in terms of subcomponents and connections. The AADL core language is designed to describe static architectures with operational modes for their components. However, it can be extended to associate additional information to the architecture. AADL error models are an extension intended to support (qualitative and quantitative) analyses of dependability attributes. The AADL Error Model Annex defines a sub-language to declare reusable error models within an error model annex library. The AADL architecture model serves as a skeleton for error model instances. Error model instances can be associated with components of the system and with the system itself.

**The component error models** describe the behavior of the components with which they are associated, in the presence of internal faults and recovery events, as well as in the presence of external propagations from the component's environment. Error models have two levels of description: the error model type and the error model implementation. The error model type declares a set of `error states`, `error events` (internal to the component) and `error propagations`[2] (events that propagate, from one component to other components, through the connections and bindings between components of the architecture model). Propagations have associated directions (`in` or `out` or `in out`). Error model implementations declare `transitions` between states, triggered by events and propagations declared in the error model type. Both the type and the implementation can declare `Occurrence` properties that specify the arrival rate or the occurrence probability of events and propagations. An `out` propagation occurs according to a specified `Occurrence` property when it is named in a transition and the current state is the origin of the transition. If the source state and the destination state of a transition triggered by an `out` propagation are the same, the propagation is sent out of the component but does not influence the state of the sender component. An `in` propagation occurs as a consequence of an `out` propagation from another component. Figure 1 shows an error model example.

---

[2] Error states can also model error free states, error events can also model repair events and error propagations can model all kinds of notifications.

```
         Error Model Type [simple]
error model simple
features
Error_Free: initial error state;
Failed: error state;
Fail: error event
      {Occurrence => Poisson λ};
Recover: error event
      {Occurrence => Poisson μ};
KO: in out error propagation
      {Occurrence => fixed p};
end simple;
```
```
 Error Model Implementation [simple.general]
error model implementation
      simple.general
transitions
Error_Free-[Fail] -> Failed;
Error_Free-[in KO] -> Failed;
Failed-[Recover] -> Error_Free;
Failed-[out KO] -> Failed;
end simple.general;
```
**Figure 1. Simple error model**

Error model instances can be customized to fit a particular component through the definition of `Guard` properties that control and filter propagations by means of Boolean expressions.

**The system error model** is defined as a composition of a set of concurrent finite stochastic automata corresponding to components. In the same way as the entire architecture, the system error model is described hierarchically. The state of a system that contains subcomponents can be specified as a function of its subcomponents' states (i.e., the system has a derived error model).

## 3. Overview of the modeling framework

For complex systems, the main difficulty for building a dependability model arises from dependencies between the system components. Dependencies can be of several types, identified in [11]: functional, structural or related to the recovery and maintenance strategies. Exchange of data or transfer of intermediate results from one component to another is an example of functional dependency. The fact that a thread runs on a processor induces a structural dependency between the thread and the processor. Sharing a recovery or maintenance facility between several components leads to a recovery or maintenance dependency. Functional and structural dependencies can be grouped into an architecture-based dependency class, as they are triggered by physical or logical connections between the dependent components at architectural level. Instead, recovery and maintenance dependencies are not always visible at architectural level.

A structured approach is necessary to model dependencies in a systematic way, to promote model reusability, to avoid errors in the resulting model of the system and to facilitate its validation. In our approach, the AADL dependability-oriented model is built in a progressive and iterative way. More concretely, in a first iteration, we propose to build the model of the system's components, representing their behavior in the presence of their own faults and recovery events only. The components are thus modeled as if they were isolated from their environment. In the following iterations, dependencies between basic error models are introduced progressively.

This approach is part of a complete framework that allows the generation of dependability analysis and evaluation models from AADL models. An overview of this framework is presented in Figure 2.

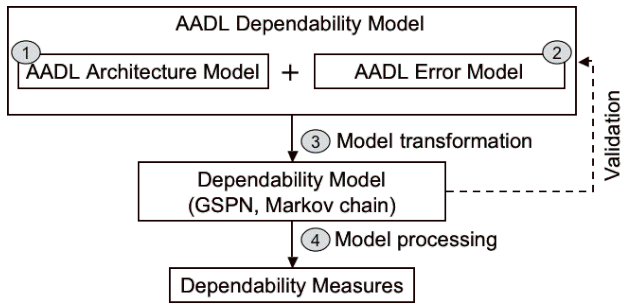

**Figure 2. Modeling framework**

The *first step* is devoted to the modeling of the application architecture in AADL (in terms of components and operational modes of these components). The AADL architecture model may be available if it has been already built for other purposes.

The *second step* concerns the specification of the application behavior in the presence of faults through AADL error models associated with components of the architecture model. The error model of the application is a composition of the set of component error models.

The architecture model and the error model of the application form the dependability-oriented AADL model, referred to as the AADL dependability model.

The *third step* aims at building an analytical dependability evaluation model, from the AADL dependability model, based on model transformation rules.

The *fourth step* is devoted to the dependability evaluation model processing that aims at evaluating quantitative measures characterizing dependability attributes. This step is entirely based on existing processing tools.

The iterative approach can be applied to the second step of the modeling framework only or to the second and third steps together. In the latter case, semantic validation based on the analytical model, after each iteration, is helpful to identify specification errors in the AADL dependability model.

Due to space limitations, we focus only on the first and second steps in this paper. A transformation from AADL to generalized stochastic Petri nets (GSPN) for dependability evaluation purposes is presented in [12].

## 4. AADL dependability model construction

To illustrate the proposed approach, the rest of this section presents successively guidelines for modeling an architecture-based dependency (structural or functional) and a recovery and maintenance dependency. More general practical aspects for building the AADL dependability model are given at the end of this section. Note that we illustrate the principles using the graphical notation for AADL composite components (system components). However, they apply to all types of components and connections.

### 4.1. Architecture-based dependency

The dependency is modeled in the error models associated with the dependent components, by specifying respectively outgoing and incoming propagations and their impact on the corresponding error model. An example is shown in Figure 3: *Component 1* sends data to *Component 2*, thus we assume that, at the error model level, the behavior of *Component 2* depends on that of *Component 1*.

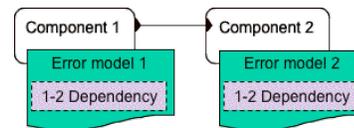

**Figure 3. Architecture-based dependency**

Instances of the same error model, shown in Figure 1, are associated both with *Component 1* and with *Component 2*. However, the AADL dependability model is asymmetric because of the unidirectional connection between *Component 1* and *Component 2*. Thus, the `out` propagation *KO* declared in the error model instance associated with *Component 2* is inactive (i.e., even if it occurs, it cannot propagate to *Component 1*).

The `out` propagation *KO* from the error model instance of *Component 1*, together with its `Occurrence` property and the AADL transition triggered by it form the "sender" part of the dependency. It means that when *Component 1* fails, it sends a propagation through the unidirectional connection. The `in` propagation *KO* from the error model instance of *Component2* together with the AADL transition triggered by it form the "receiver" part of the dependency. Thus, an incoming propagation *KO* causes the failure of the receiving component.

In real applications, architecture-based dependencies usually require using more advanced propagation controlling and filtering through `Guard` properties. In particular, Boolean expressions can be defined to specify the consequences of a set of propagations occurring in a set of sender components on a receiver component.

### 4.2. Recovery and maintenance dependency

Recovery and maintenance dependencies need to be described when recovery and maintenance facilities are shared between components or when the maintenance activity of some components has to be carried out according to a given order or a specified strategy (i.e., a thread can be restarted only if another thread is running).

Components that are not dependent at architectural level may become dependent due to the recovery and maintenance strategy. Thus, the AADL dependability model might need some adjustments to support the description of dependencies related to the maintenance strategy. As error models interact only via propagations through architectural features (i.e., connections, bindings), the recovery and maintenance dependency between components' error models must be supported by the architecture model. Thus, besides the architecture components, we may need to model (at architectural level) a component allowing to describe the recovery and maintenance strategy. Figure 4-a shows an example of AADL dependability model. In this architecture, *Component 3* and *Component 4* do not interact at the architecture level. However, if we assume that they share a recovery and maintenance facility, the recovery and maintenance strategy has to be taken into account in the error model of the application. Thus, it is necessary to represent the recovery and maintenance facility at the architectural level, as shown in Figure 4-b in order to model explicitly the dependency between *Components 3* and *Component 4*.

Also, the error models of dependent components with regards to the recovery and maintenance strategy might need some adjustments. For example, to represent the fact that *Component 3* can only restart if *Component 4* is running, one needs to distinguish between a failed state of *Component 3* and a failed state where *Component 3* is allowed to restart.

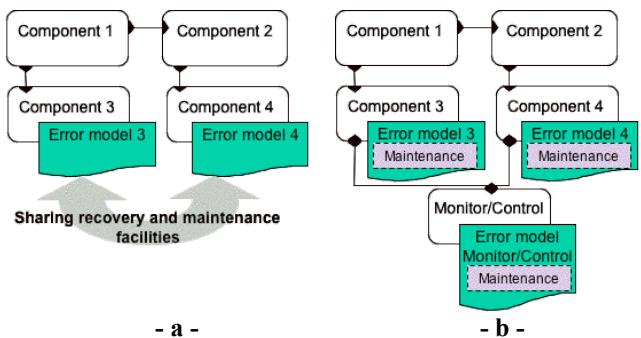

**Figure 4. Maintenance dependency**

### 4.3. Practical aspects

The order for modeling dependencies does not impact the final AADL dependability model. However, it may impact the reusability of parts of the model. Thus, the order may be chosen according to the context of the targeted analysis. For example, if the analysis is meant to help the user to choose the best-adapted structure for a system whose functions are completely defined, it may be convenient to introduce first functional dependencies between components and then structural dependencies, as the model corresponding to functional dependencies is to be reused. Generally, recovery and maintenance dependencies are modeled at the end, as one important aim of the dependability evaluation is to find the best-suited recovery and maintenance strategies for an application. Recovery and maintenance dependencies may have an impact on the system's structure.

Not all the details of the architecture model are necessary for the AADL dependability model. Only components that have associated error models and all connections and bindings between them are necessary. This allows a designer to evaluate dependability measures at different stages in the development cycle by moving from a lower fidelity AADL dependability model to a detailed one. In some cases, not all components having associated error models are part of the AADL dependability model. The AADL Error Model Annex offers two useful abstraction options for error models of components composed of subcomponents:

– The first option is to declare an abstract error model for a system component. In this case, the corresponding component is seen as a black box (i.e., the detailed subcomponents' error models are not part of the AADL dependability model). This option is useful to abstract away modeling details in case an architecture model with too detailed error models associated with components does exist for other purposes. Issues linked to the relationship between abstract and concrete error models have been mentioned in [13].

– The second option is to define the state of a system component as a function of its subcomponents' states. This option can be used to specify state classes for the overall application. These classes are useful in the evaluation of measures. If the user wishes to evaluate reliability or availability, it is necessary to specify the system states that are to be considered as failed states. If in addition, the user wishes to evaluate safety, it is necessary to specify the system states that are considered as catastrophic.

## 5. Example

In this section we illustrate our modeling approach on a small software architecture representing a process whose functional role is to compute a result. The computation is divided in three sub computations, each of them being performed by a thread. The thread *Compute2* uses the result obtained by the thread *Compute1* and the thread *Compute3* uses the result obtained by the thread *Compute2* to compute the result expected from the process. The three threads are connected through data connections according to the pipe and filter architectural style [14]. Due to space limitations, we only take into account two dependencies:

- An architecture-based dependency between the computing threads: a failure in one of the computing threads may cause the failure of the following thread (with a probability *p*). In some cases, cascading failures can occur.
- A recovery dependency: *Compute3* can only recover if *Compute1* and *Compute2* are error free. We assume that *Compute2* can recover if *Compute1* is not error free.

The AADL dependability model of this application is shown in Figure 5 using the AADL graphical notation.

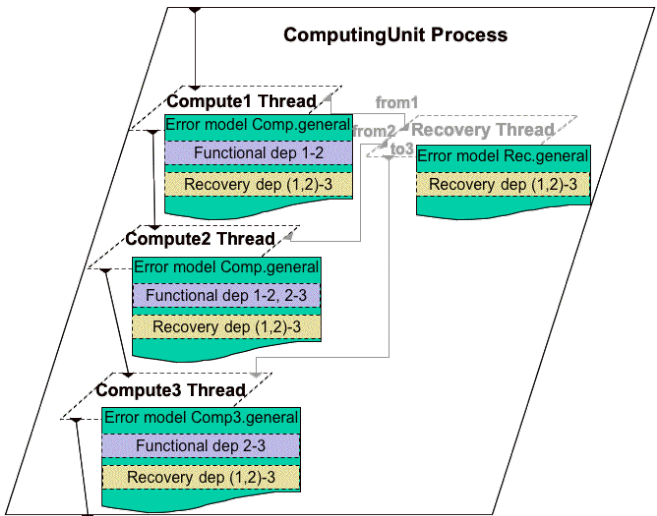

**Figure 5. AADL dependability model**

The AADL dependability model of this application is built in three iterations. The computing threads' behavior in the presence of their own fault and recovery events is represented in the first iteration. The propagation *KO* together with corresponding transitions are added in a second iteration to represent the architecture-based dependency. The thread *Compute1* can have an impact on *Compute2* and *Compute2* can have an impact on *Compute3*. We remind that the opposite is not possible, as the connections between threads are unidirectional. The recovery dependency is modeled in the third iteration. It requires the existence of a *Recovery* thread in the architecture model (see light grey part of Figure 5). Its role is to send (through the `out` port *to3*) a *RecoverAuthorize* propagation to *Compute3* if *Compute1* and *Compute2* are error free.

Figure 6-a shows the error model *Comp.general* associated with threads *Compute1* and *Compute2*. Figure 6-b shows the error model *Comp3.general* associated with the threads *Compute3*. The three iterations are highlighted. Each line tagged with a (+) sign is added to the error model corresponding to the previous iteration while each line tagged with a (-) sign is removed from it during the current iteration. The first and second iterations are the same for all three computing threads. In the third iteration, it is necessary to distinguish between a failed state and a failed state from which *Compute3* is authorized to restart. This leads to removing a transition declared in the first iteration, and adding a state (*CanRecover*) and two transitions linking it to the state machine.

Figure 7 shows the `Guard_Out` property applied to port *to3* of the *Recovery* thread in the third iteration. This property specifies that a *RecoverAuthorize* propagation is sent to *Compute3* through port *to3* when *OK* propagations are received through ports *in1* and *in2* (meaning that *Compute1* and *Compute2* are error free). The *Recovery* thread has an associated error model that is not shown here. It declares `in` and `out` propagations used in the `Guard_Out` property.

The main idea of this method is to verify and validate the model at each iteration. If a problem arises during iteration *i*, only the part of the current AADL dependability model corresponding to iteration *i* is questioned. Thus, the validation process is facilitated especially in the context of complex systems.

## 6. Conclusion

This paper presented an iterative approach for system dependability modeling using AADL. This approach is meant to ease the task of analyzing dependability characteristics and evaluating dependability measures for the AADL users community. Our approach assists the user in the structured construction of the AADL dependability model (i.e., architecture model and dependability-related information). To support and trace model evolution, this approach proposes that the user builds the model iteratively. Components' behaviors in the presence of faults are modeled in the first iteration as if they were isolated. Then, each iteration introduces a new dependency between system components. Error models representing the behavior of several types of system components and several types of dependencies may be placed in a library and then instantiated to minimize the modeling effort and maximize the reusability of models.

The OSATE toolset is able to support our modeling approach. It also allows choosing component models and error models from libraries. For the sake of illustration, we used simple examples in this paper. We have already applied the iterative modeling approach to a system with multiple dependencies in [12] and we plan to validate it against other complex case studies.

```
       Error Model Type [Comp]
error model Comp
features
-- iteration 1
(+) Error_Free: initial error state;
(+) Failed: error state;
(+) Fail: error event
(+)     {Occurrence => Poisson λ};
(+) Recover: error event
(+)     {Occurrence => Poisson μ};
-- iteration 2
(+) KO: in out error propagation
(+)     {Occurrence => fixed p};
-- iteration 3
(+) OK: out error propagation
(+)     {Occurrence => fixed 1};
end Comp;
```
```
Error Model Implementation [Comp.general]
error model implementation Comp.general
transitions
-- iteration 1
(+) Error_Free-[Fail]->Failed;
(+) Failed-[Recover]->Error_Free;
-- iteration 2
(+) Error_Free-[in KO]->Failed;
(+) Failed-[out KO]->Failed;
-- iteration 3
(+) Error_Free-[out OK]->Error_Free;
end Comp.general;
```
**a: Error Model for Compute1 and Compute2**

```
       Error Model Type [Comp3]
error model Comp3
features
-- iteration 1
(+) Error_Free: initial error state;
(+) Failed: error state;
(+) Fail: error event
(+)     {Occurrence => Poisson λ};
(+) Recover: error event
(+)     {Occurrence => Poisson μ};
-- iteration 2
(+) KO: in out error propagation
(+)     {Occurrence => fixed p};
-- iteration 3
(+) CanRecover: error state;
(+) OK: in error propagation;
end Comp3;
```
```
Error Model Implementation [Comp3.general]
error model implementation Comp3.general
transitions
-- iteration 1
(+) Error_Free-[Fail]->Failed;
(+) Failed-[Recover]->Error_Free;
-- iteration 2
(+) Error_Free-[in KO]->Failed;
(+) Failed-[out KO]->Failed;
-- iteration 3
(-) Failed-[Recover]->Error_Free;
(+) Failed-[RecoverAuthorize]->CanRecover;
(+) CanRecover-[Recover]->Error_Free;
end Comp3.general;
```
**b: Error Model for Compute3**

**Figure 6. Error model for Compute1 / Compute2**

```
     Guard_Out [port Recovery.to3]
-- iteration 3
(+) Guard_Out =>
(+)     RecoverAuthorize when
(+)     (from1[OK] and from2[OK])
(+)     mask when others
(+) applies to to3;
```
**Figure 7. Guard_Out property (port Recovery.to3)**

## Acknowledgements

This work is partially supported by 1) the European Commission (European integrated project ASSERT No. IST 004033 and network of excellence ReSIST No. IST 026764). and 2) the European Social Fund.